# Optimized energy utilization in small and large commercial loads and residential areas

Hayder O. Alwan, Hamidreza Sadeghian, Sherif Abdelwahed

*Abstract* – In smart grid, the demand side management (DSM) techniques need to be designed to process a large number of controllable loads of several types. In this paper, we proposed a framework to study the demand side management in smart grid which contains a variety of loads in two service areas, one with multiple residential households, and one bus with commercial customers. Specifically, each household may have renewable generation as well as interruptible and uninterruptible appliances to make individual scheduling to optimize the electric energy cost by making the best time of the electricity usage according to the day ahead forecast of electricity prices. A high load bus represents a commercial area employed to demonstrate the impact of high load at any bus on voltage profile, power loss, and load flow condition, and to show the performance of the proposed DSM for large number of appliance. Using the developed simulation model, we examine the performance of the proposed DSM and study their impact on the distribution network operation and renewable generation, overall voltage deviation, real power loss, and possible problems such as reverse power flows, voltage rise have examined and compared, these problems can easily be seen at the commercial load bus.

*Index-Term-- Demand* Side Management, Distribution Generation, Rooftop Photovoltaic, Energy Utilization, Optimization algorithm.

## I. INTRODUCTION

The benefits of DSM include financial and system reliability, among others. Financial benefits are gained in the bill savings and incentive payments earned by customers that adjust their electricity demand in response to time-varying electricity rates or incentive-based programs [1]. Reliability benefits are the operational security and adequacy savings that result because DSM lowers the likelihood and consequences of forced outages that impose financial costs and inconvenience on customers [2]. In the United States, many DSM programs are widely implemented by commercial and industrial customers. These are mainly interruptible load, direct load control, real-time pricing and time-of-use programs [3]. On the other hand, very few DSM programs are in use today for residential customers. Authors in [4] put forward a scheduling approach of operation and energy consumption of various electrical appliances in a grid connected smart home system. Reference [5]and [6] introduced an algorithm that simulated residential load shifting under time of use (TOU) regimes using previously generated profile data to model realistic demand response behavior. Reference [7] shows the economic benefits of DSM on the agriculture and industrial sectors. Majority of DSM application have focused on large commercial loads, these loads have a large amount of demand to make a considerable contribution to the stability of the grid [8].

In this paper develops an optimization model for a multiple residential households and two size of commercial loads with a rooftop PV installation. This algorithm has the ability to take into consideration the evolution of the system performance in terms of operation parameter such as voltage fluctuation, power loss of the entire system, and the PV utilization efficiency while optimizing the electricity cost. In addition the proposed algorithm can handle large number of controllable appliances in two types of loads residential and commercial taking into consideration the fact that certain appliances may have higher priority over other appliances so that these appliances may shifted to the suitable time according to their importance. In the simulation process the algorithm classifies the commercial appliances into three categories: high, med and low. The appliances in each category subjected to different penalty prices according to the importance of the appliance. The rest of this paper is organized as follows. Section II presents the system modeling for the proposed study, section III defines the optimization model of decentralized DSM with rooftop PV at each customer. Section IV provides numerical simulation results and analyzes the impacts of different DSM schemes on grid operation. Finally, section V is the conclusion.

## II. SYTEM MODELING

This section describes a system model for the proposed DSM in a single radial distribution network with thirty buses to demonstrate the effectiveness of the proposed approach, the DSM strategy is tested on two different areas, each with different type of customers; residential and commercial. Each area has different type of controllable appliances.

### A. Residential Community area

As depicted in Fig. 1, bus 1 represents a substation while the rest of the buses represent the simulated residential community with up to 29 households with a more diverse combination of appliances. Each smart home will optimize individually using the proposed decentralized DSM algorithm its appliance operation schedule to save electricity costs according to the day-ahead residential and commercial time of use pricing (TOU) as shown in fig .4 and rooftop PV (if available).

### B. Commercial Area

The devices subjected to load control in the commercial area (delivered by Bus No.17) have consumption ratings which are slightly higher than those in the. Figs 4 and 5. Show the curves for the one commercial activity, also the curve show that there is one peak load occurs during the period from 9:00 AM to 5:00 PM. Each appliance is modeled using four parameters $s_a, f_a, r_a$ and $D_a$, where $[s_a, f_a]$ defines the allowable operating time during which the appliance $a$ may be switched on, $r_a$ and $D_a$ denote the power rating and the total number of operating time slots as requested, respectively.

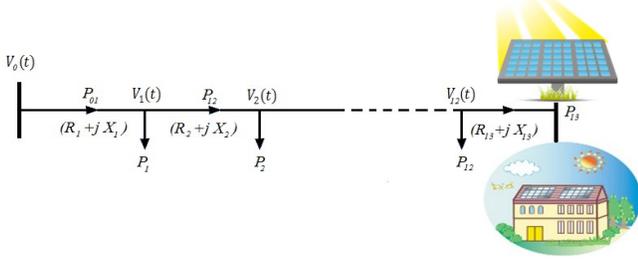

Fig.1. Distribution network of households

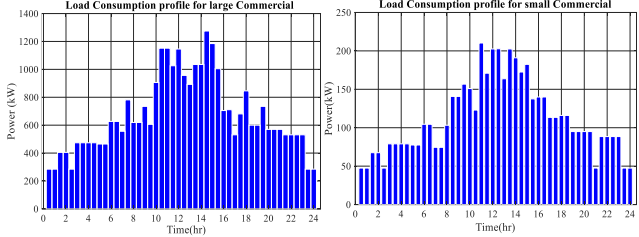

Fig.2 Large Commercial Load    Fig.3 Small Commercial Load

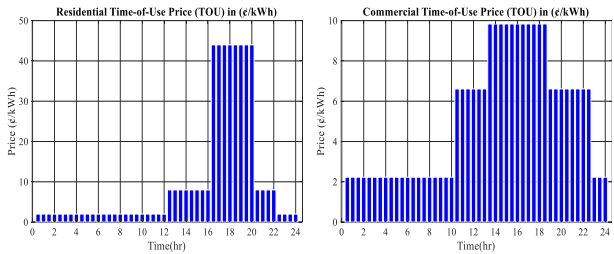

Fig .4 Time-of –Use (TOU) Price for Residential and Commercial loads

### III. MATHEMATICAL FORMULATION

The DSM optimization for each households can be defined as below:

$$\min_{[u_{a,m}(t)]} C_e^m + C_p^m \quad (1)$$

Subject to:
$$C_e^m = 0.5 \times \sum_t^T P_{load}^m(t) \times \pi_e(t) \quad (2)$$
$$C_p^m = 0.5 \times \sum_{a,m=1}^A \pi_p \cdot r_{a,m} \cdot \Delta T_{a,m} \quad (3)$$
$$P_{load}^m(t) = \max((\sum_{a,m=1}^{A_m} r_{a,m} \times u_{a,m}(t) - \alpha. P_{pv}^m(t)), 0) \quad (4)$$
$$\sum_{a,m=1}^{A_m} r_{a,m} \times u_{a,m}(t) \leq MD^m \quad \forall a \in \{1 \text{ to } A_m\} \quad (5)$$
$$\sum_{t=1}^T u_{a,m}(t) = D_{a,m} \quad \forall a \in \{1 \text{ to } A_m\} \quad (6)$$
$$u_{a,m}(t) = 0 \quad \forall t < s_{a,m} \text{ or } \forall t > f_{a,m} \quad (7)$$
$$\Delta T_{a,m} = 1^T. |t_{a,m}^{st_{new}} - t_{a,m}^{st_{old}}| \quad \forall a \in \{1 \text{ to } A_m\} \quad (8)$$
$$t_{a,m}^{st_{new}} = [t|u_{a,m}^{new}(t) = 1]_{1 \times D_{a,m}} \quad \forall a \in \{1 \text{ to } A_m\} \quad (9)$$
$$t_{a,m}^{st_{old}} = [t|u_{a,m}^{old}(t) = 1]_{1 \times D_{a,m}} \quad \forall a \in \{1 \text{ to } A_m\} \quad (10)$$

Where $m$ is the housholds index, $u_{a,m}$ Represents a binary status of appliance $a$; 0 = *off*, 1 = on at Household $m$, with following format:
$$[u_{a,m}(t)]_{A \times T} = [u_1^1, u_1^2, \ldots, u_1^T; u_2^1, u_2^2, \ldots u_2^T; \ldots; u_A^1, u_A^2, \ldots, u_A^T] \quad (1)$$
Where T is total number of time slots, T=48, and t is the index of the time slots. Each appliance is modeled using four parameters $s_a, f_a, r_a$ and $D_a$, where $[s_{am}, f_{a,m}]$ defines the allowable operating time during which the appliance $a$ in the household $m$ may be switched on, $r_{a,m}$ and $D_{a,m}$ denote the power rating and the total number of operating time slots as requested in the household $m$ respectively. Where (2)-(3) define the electricity cost and the penalty cost; Eq. (5) to avoid negative electricity cost. In our proposed model, we have assumed that surplus PV generation will be injected into the distribution network without reward, so the total electricity cost within each time slot should be no less than zero. α in Eq. (5) is a binary parameter stands for status of PV installation at DSM household. Constraint (6) indicates the Maximum Demand (MD) that the aggregate appliance power of each household cannot exceed at any time. This specified upper limit is to prevent super-high-power demand peak even during the hours when day-ahead electricity price is low because the utilities do not want to have "new" peak created by the DSM load-shifting or because the distribution feeders have capacity constraints. Constraint (7) and (8) indicate the total operation duration and the allowable turn-on time of an appliance. Constraints (9)indicates that for the base load the time difference between original and new starting points equal to zero Constraints (10)-(12) specify the original and the new starting point, $t_{a,m}^{st_{old}}$ and $t_{a,m}^{st_{new}}$ in the household $m$ respectively, to capture the duration of time-shifting for flexible appliances.

### IV. NUMERICAL SIMULATION RESULTS

#### A. DSM Residential Area

The red sold line illustrates the original load profiles for each household. The performance of the implemented DSM algorithm was tested on two configurations. Each of the 29 households has rooftop PV. The result is a significant improvement in the daily load consumption pattern and a reduction of total electricity cost. According to Fig.4. The cost of households 1, 2, 3, and 4 were $4.08, $6.41, $3.88, and $3.65 per day, respectively. Comparing those costs with the original costs, household 1 sees a reduction of 76%. In households 2, 3, and 4, the cost reduction is 70%, 70.23%, and 66%, respectively. Most peak loads during high-priced hours

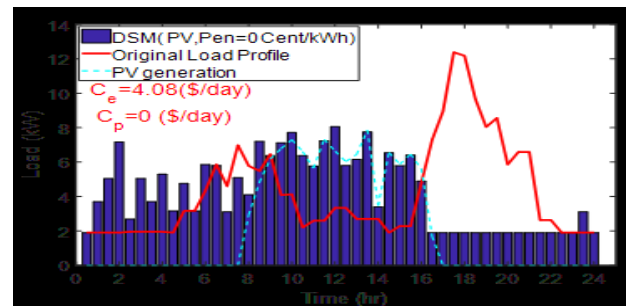

moved to off-peak periods, except for the non-flexible appliances. Also, with local PV generation, during the time slots with PV availability, appliance operation is free

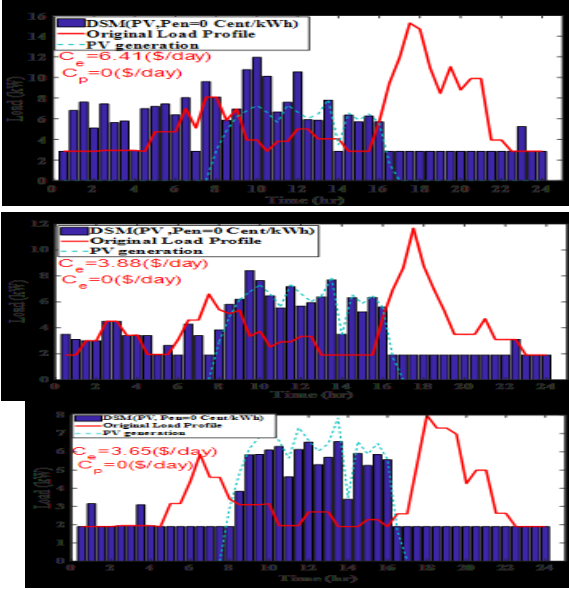

Fig.5 Load profile of smart household #1 DSM four - different conditions

### B. Voltage Fluctuations

To check if there is any voltage violation, Fig. 6 clarify the voltage deviation of the Bus No. 17 for the small commercial load, as we can see that there was a slight difference between the voltage profile in case of considering DSM with PV, the voltage increased between the 12:30 pm and 4:00 pm as most of the demanded load covered by the generated PV power. In other word, the voltage rise due to the reverse power flow can be suppressed by reducing the amount of active power produced by PV. In this work, Voltage rise occurs when the load demand is low and the PV generation in its max level. The possible solution for voltage rise can be done during our simulation by either reduce the network resistance. Or by reduce the PV penetration Level. As we can see the reduction in the PV penetration level to the half caused slight reduction in voltage level. Maximum and minimum feeder voltages were recorded for each simulation, and simulations were continued at increasing PV levels as depicted in Fig.6.

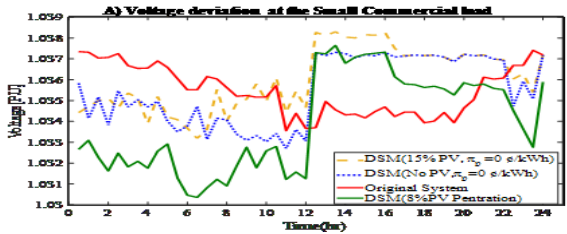

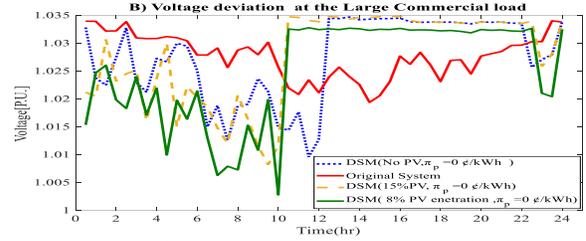

Fig.6 Voltage profile at commercial load

### C. Real Power loss

From fig. 7, it is clearly that, when injecting PV generated power, we can reduce the distribution line losses, as we can see during the time 8:00-12:30 there is a reduction in power loss. The overall reduction in power loss was 26%. This is because PV generated power output supplies a portion of real and reactive power to the load. As PV systems generate power locally to meet the customer demands, appropriate size can drastically reduce power losses in the system. As we can see in fig. 7 (A) black dotted line, the case of DSM (No PV, $\pi_p$ =0 ¢/kWh) there is an increased in the power loss between 8 Am and 12:00 PM, the load demand at this period of time is high and the current drawn will be high. The blue line represents the power loss in the line feeder in the presence of PV generation, as we can see in fig. 8 for the period from 8:00 AM to 12:00 the power loss has significant reduction, due to the fact that the residential loads and the commercial bus are using the free power generated by PV. Table I show the impact increasing penalty factor in residential and commercial load on the PV utilization efficiency, it shows that utilization efficiency decreased for residential and we explain that in this paper, see "Individual Households DSM", while it decreased in commercial increased as PV highest power generated at the period of peak commercial load see Figs. 2 and 3. Generally, the total power losses in a distribution system calculated using:

$P_{Loss} = \sum_{i=1}^{n_{br}} |I_i|^2 \, r_i$ , Where $n_{br}$ the total number of branches is in the system $|I_i|$ is the magnitude of the current flow in branch $i$, $r_i$ is the resistance of branch $i$.

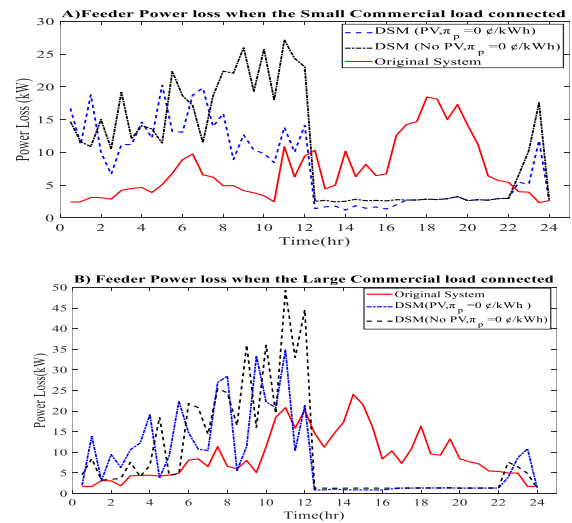

Fig. 7 Overall feeder power Loss when A) Small commercial load connected B) Large commercial load connected

Table. I PV utilization efficiency at different penalty price $\pi_p = 0[¢/kWh]$

A. *DSM in the Commercial Area*

The results obtained for the commercial area are given in Fig. 9, the comparison of the original load profiles shown in red solid line, as we can see there is one demand peaks around 9:00 to 4:00 and the new distribution load profile after applying DSM illustrated by the blue bar. We divided the appliance in the commercial area [as in Appendix B]. Fig.10-a Illustrates the cost saving of the shifted appliances and the corresponding penalty cost for $\pi_p = 0,1,3$ ¢/kWh which reported the penalty prices applied for Low, Med, and high critical respectively. The obtained results show that one high critical appliance participated in the DSM with one time slots shifts, while five appliances under Med-critical categories participated in the DSM and the number of the shifted slots increases, it's worth noting that the corresponding cost saving depends not only on the reduced electricity price caused by the shifted $\Delta T_a$ slots but also on the power rate $r_a$ of the appliance. Lastly for Low-critical we see that number of the appliances participating in DSM increase to nine with much higher cost saving and higher time shifted slots $\Delta T_a$, also we see from Fig.8-(a) that the low-critical appliances have zero penalty cost as $\pi_p = 0$ ¢/kWh.

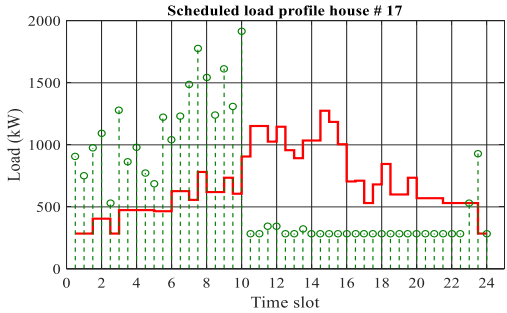

Fig.7 DSM results for the commercial at $\pi_p = 0$ ¢/kWh

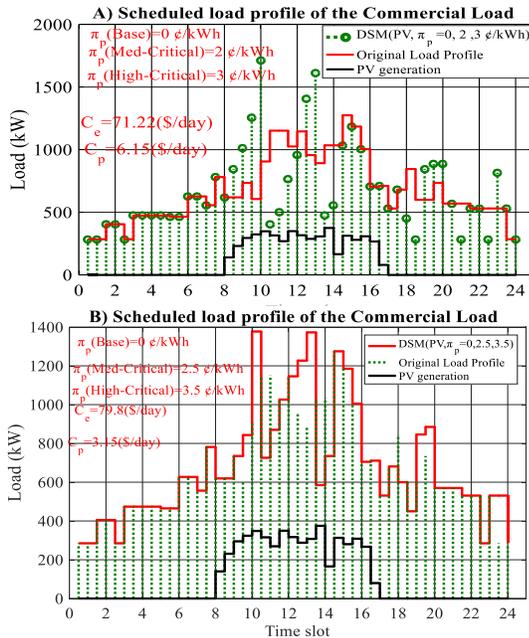

Fig.8. A and B Represnts DSM Results for commercial at Different $\pi_p$

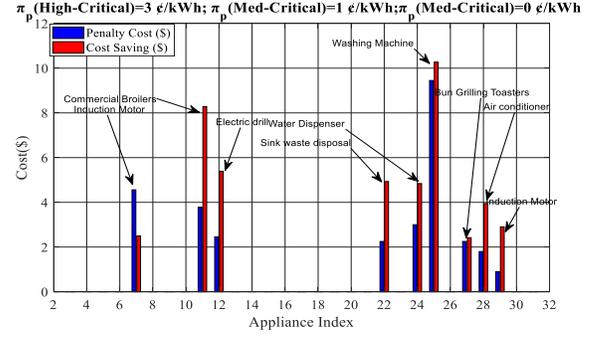

Fig.11. Cost saving and Penalty Cost for Commercial DSM

Table II. Feeder Power Loss in kW for Different Participation Level

|  | %100 DSM | 50%DSM | %25 DSM |
|---|---|---|---|
| **DSM with PV** | 136.5 | 153.6 | 172.5 |
| **DSM without PV** | 202.63 | 199.97 | 197.33 |

V. Conclusion

For radial grid layout comprising 30 buses, we assessed (1) the reduction in the operating cost, (2) PV utilization efficiency, (3) real rower loss, and (4) Voltage Fluctuation. The electricity cost of the residential area show reduction of 37.1%, while the commercial area show reduction of 50.3%, although the customers in commercial building less willing to change their consumption patterns, due to high power rate of the appliances in the commercial load. The he reduction in power loss in the commercial area was higher than the reduction of power loss in residential load, as the peak load in the original commercial load profile occur at the time of the peak PV generation period Lastly, the results obtain from voltage profile show that when the value $\pi_p$ increased the voltage less flattering and more fluctuated, also fewer appliances can shift to the time of the PV power generated, in other word, apply DSM without high penalty for load-shifting will, reduce the voltage rise, encourage renewable energy consumption, and avoid the overvoltage might occurs due to high penetration level.

| PV Utilization Efficiency | | | |
|---|---|---|---|
| Penalty Price $\pi_p = $ [¢/kWh] | Residential Area | Penalty Price $\pi_p = $ [¢/kWh] | Commercial Area |
| 0 | 98% | 0 | 96% |
| 5 | 68.3% | 2 | 100% |
| 10 | 62.5% | 3 | 100% |
| 20 | 58.4% | 5 | 100% |